\documentclass[smallextended]{svjour3}

\RequirePackage[T1]{fontenc}

\smartqed  % flush right qed marks, e.g. at end of proof

\RequirePackage{graphicx}
\RequirePackage{mathptmx}      % use Times fonts if available on your TeX system
\RequirePackage{flushend}
\RequirePackage[numbers,sort&compress]{natbib}
\RequirePackage[colorlinks,citecolor=blue,urlcolor=blue,linkcolor=blue]{hyperref}
\usepackage{amsmath}
\usepackage{amssymb}

\journalname{Gen. Rel. Grav.}

\def\to{\rightarrow}

\def\al{\alpha} \def\be{\beta} \def\ga{\gamma} \def\de{\delta}
\def\ep{\epsilon}   
\def\th{\theta}   \def\ka{\kappa}
\def\la{\lambda}

\def\om{\omega} \def\Ga{\Gamma} \def\De{\Delta} 
   
 \def\Om{\Omega} \def\mn{{\mu\nu}} \def\cl{{\cal L}}

 \def\frac#1#2{{\textstyle{{#1}\over
{#2}}}} 
\def\lsim{\mathrel{\rlap{\lower4pt\hbox{\hskip1pt$\sim$}}
\raise1pt\hbox{$<$}}}
\def\gsim{\mathrel{\rlap{\lower4pt\hbox{\hskip1pt$\sim$}}
\raise1pt\hbox{$>$}}} \def\sqr#1#2{{\vcenter{\vbox{\hrule height.#2pt
\hbox{\vrule width.#2pt height#1pt \kern#1pt \vrule width.#2pt} \hrule
height.#2pt}}}}
\def\square{\mathchoice\sqr66\sqr66\sqr{2.1}3\sqr{1.5}3}
\def\beq{\begin{equation}} \def\eeq{\end{equation}}
\def\beqa{\begin{eqnarray}} \def\eeqa{\end{eqnarray}}
\def\eq#1{Eq. (\ref{#1})}

\begin{document}

\title{Homogeneous spherically symmetric bodies with a non-minimal coupling between curvature and matter}
\subtitle{The choice of the Lagrangian density for matter}
\titlerunning{Homogeneous spherically symmetric bodies with a non-minimal coupling}

\author{Orfeu Bertolami \and Jorge P\'aramos}

\institute{Centro de F\'isica do Porto and Departamento de F\'isica e Astronomia,\\Faculdade de Ci\^encias da Universidade do Porto,\\Rua do Campo Alegre 687, 4169-007 , Porto, Portugal\\ \email{orfeu.bertolami@fc.up.pt \and jorge.paramos@fc.up.pt}}

\date{Received: \today / Accepted: }

\maketitle

\begin{abstract}

In this work we study how a non-minimal coupling between matter and gravity can modify the structure of a homogeneous spherical body. The physical relevance of the adopted Lagrangian density is ascertained, with results obtained for two different choices of the latter.
\end{abstract}

\section{Introduction}

Despite its great experimental success (see {\it e.g.} Refs. \cite{Will2006,OBJP2012}), General Relativity (GR) is not the most encompassing way to couple matter with curvature. Indeed, matter and curvature can be coupled, for instance, in a non-minimal way \cite{model}, extending the well-known class of $f(R)$ theories \cite{Felice,BirdsEye,SotiriouReview}. This nonminimal coupling can have a bearing on the dark matter \cite{model,DM1,DM2} and dark energy \cite{DE1,DE2,DE3,DE4} problems, impact the well-known energy conditions \cite{EC}, affect the Yukawa potential addition prompted by $f(R)$ theories in Solar System tests of gravity \cite{naf,NCB}, modify the Layzer-Irvine equation of virial equilibrium \cite{Bertolami:2014hpa} and can give rise to wormhole and time machines \cite{WHTM}. In Ref. \cite{Martins}, several phenomenological aspects of the dynamics of perfect fluids non-minimally coupled to curvature were addressed ---  in particular, the scenario of an axisymmetric dust distribution with constant density.

Another interesting issue that arises in the context of gravity theories with a non-minimal coupling between curvature and matter is the fact that the Lagrangian degeneracy in the description of a perfect fluid, encountered in GR \cite{1968RSPSA.305....1S,Schutz,Brown} is lifted \cite{Lagrangian}: indeed, since this quantity explicitly appears in the modified equations of motion, two Lagrangian densities leading to the same energy-momentum tensor have different dynamical implications, whereas in GR they are physically indistinguishable.

In minimally coupled $f(R)$ theories, static spherical symmetry was studied in Refs. \cite{fRspherical1,fRspherical2,fRspherical3,fRspherical4}, with gravitational collapse explored in Refs. \cite{Cembranos,Sharif,Ghosh}.

In what follows we shall examine the role that two possible descriptions of a perfect fluid ({\it i.e.} two different choices of the Lagrangian density) have on the structure of a spherically symmetric gravitational body. Previous works on spherical solutions with a nonminimal coupled scenario include a polytropic equation of state \cite{solar}, constant curvature solutions \cite{DE2}, the embedding of a spherical body on the background cosmological fluid \cite{DE3}, the collapse of a homogeneous body \cite{collapse} and black hole solutions \cite{Bertolami:2014pga}. This does not contradict the previous work where it is argued that a more suitable choice for the Lagrangian density of a perfect fluid is $\cl = -\rho$ \cite{Lagrangian} --- but aims to show that, under adequate circumstances, the adopted Lagrangian density is not all that crucial in determining the observable implications of the non-minimal coupling between matter and curvature.

\section{The model}

One considers a model that exhibits a non-minimal coupling (NMC) between geometry and matter, as expressed in the action functional \cite{model},

\beq S = \int \left[ \ka f_1(R) + f_2(R) \mathcal{L} \right] \sqrt{-g} d^4 x~~. \label{model}\eeq

\noindent where $f_i((R)$ ($i=1,2$) are arbitrary functions of the scalar curvature, $R$, $g$ is the determinant of the metric and $\ka = c^4/16\pi G$.

Variation with respect to the metric yields the modified field equations,

\beq \label{field0} \left( \ka F_1 + F_2 \cl \right) G_\mn =  {1 \over 2} f_2 T_\mn + \De_\mn \left( \ka F_1 + F_2 \cl \right) + {1 \over 2} g_\mn \left[ \ka(f_1 - F_1 R) - F_2 R \cl \right] ~~, \eeq

\noindent with $F_i \equiv df_i/dR$ and $\De_\mn = \nabla_\mu \nabla_\nu - g_\mn \square$. As expected, GR is recovered by setting $f_1(R) = R $ and $f_2(R) = 1$.

The trace of \eq{field0} reads

\beq \left( \ka F_1  + F_2  \cl \right) R  =   {1 \over 2} f_2 T -3 \square \left( \ka F_1 + F_2 \cl \right) + 2 \ka f_1 ~~. \label{trace0}\eeq

\subsection{Perfect fluid description}
\label{perfectfluid}

In this work, matter will be assumed to behave as a perfect fluid, {\it i.e.} a fluid with no viscosity, vorticity or heat conductivity, and described by its four-velocity $u_\al$ and a set of thermodynamical variables: the particle number density $n$, energy density $\rho$, pressure $p$, temperature $T$ and entropy density $s$. These are related by the usual laws of thermodynamics, and constrained by the physical requirements of particle number conservation $(nu^\mu)_{;\mu} = 0$, no exchange of entropy with neighbouring flow lines, $(nsu^\mu)_{;\mu} = 0$ and fixed fluid flow lines at the boundaries of spacetime; furthermore, the four velocity is normalized through $u^\mu u_\nu = -1 $ and obeys $ u^\mu \nabla_\nu u_\mu = 0$.

When looking for a solution for the field equations, the well-known energy-momentum tensor of a perfect fluid is thoroughly used,

\beq T_\mn = (\rho + p)u_\mu u_\nu + p g_\mn ~~, \label{EMtensor}  \eeq

\noindent and no particular attention is paid to the underlying Lagrangian density of a perfect fluid. However, it is clear that the above form for the energy-momentum tensor should arise from the variation of the action, 

\beq T_\mn = -{2 \over \sqrt{-g}} {\de \left(\sqrt{-g} \cl \right) \over \de g^\mn } ~~. \eeq

\noindent While a clear knowledge of $\cl$ is not relevant for most applications found in GR, it is paramount in the present case, where the NMC between matter and curvature leads to an explicit dependence of the field Eqs. (\ref{field0}), as mentioned before.

The identification $\cl = p$ was first advanced in Ref. \cite{1968RSPSA.305....1S}, followed by a relativistic generalization \cite{Schutz}. Much later, Ref. \cite{Brown} showed that this choice is equivalent to $\cl = -\rho$, complemented by a suitable set of thermodynamical potentials and Lagrange multipliers that enforce the aforementioned constraints $(nu^\mu)_{;\mu} = 0$ and $(nsu^\mu)_{;\mu} = 0$.

These different forms for the Lagrangian densities are found to be equivalent on-shell, {\it i.e.} by substituting the field equations derived from the matter action back into the action functional and reading the resulting Lagrangian density. As such, one finds that the Lagrangian densities $\cl_1 = p $ and $\cl_2 = na $ (with $a(n,T) = \rho(n)/n - sT$ the free energy) are the ``on-shell'' equivalent to the ``bare'', original functional $\cl = -\rho$.

Similarly, formally distinct expressions for the energy-momentum tensor can be derived, and shown to be dynamically equivalent: indeed, the Lagrange multiplier method of Ref. \cite{Brown} yields

\beq T_\mn = \rho u_\mu u_\nu +  \left( n {\partial \rho \over \partial n} - \rho \right) (g_\mn + u_\mu u_\nu) ~~,\eeq

\noindent which yields \eq{EMtensor} if one defines the pressure as 

\beq p \equiv n {\partial \rho \over \partial n} - \rho~~, \label{pressdef} \eeq

\noindent and requires an equation of state (EOS) $\rho = \rho(n)$. For non-relativistic motion, $\rho \sim n$ and one thus finds the usual description of dust matter, {\it i.e.} $p=0 $ and $T_\mn = \rho u_\mu u_\nu$. An alternative form for the energy-momentum tensor reads \cite{lrr-2007-1}

\beq T_\mn = p g_\mn + n \mu u_\mu u_\nu ~~, \eeq

\noindent where one uses the chemical potential $\mu = d\rho / dn$; this form is motivated by the Hamiltonian formulation of a perfect fluid, as the momentum $\mu u_\mu$ is canonically conjugate to the particle number density current $n u^\mu$. Again, the definition \eq{pressdef} establishes the equivalence between this form and the commonly used \eq{EMtensor}.  

When generalizing to a NMC scenario, Ref. \cite{Lagrangian} found that the above discussion still holds, as the effect of the NMC is shifted to the relation between the thermodynamical variables and Lagrange multipliers. Although in that study it is argued that $\cl = -\rho$ is the appropriate form to insert into the modified field Eqs. (\ref{field0}), it is relevant to ascertain to what extent do different choices of $\cl$ have an impact in particular scenarios --- such as the structure of a static, spherically symmetric body. 

\subsection{Non-conservation of the energy-momentum tensor}
\label{conservation}

Before proceeding with the discussion of a static, spherically symmetric spacetime, one first discusses the possibility of breaking the covariant conservation of the energy-momentum tensor and, due to it, the Weak Equivalence Principle (WEP).

Resorting to the Bianchi identities, one concludes that the energy-momentum tensor of matter may not be (covariantly) conserved, since

\beq \nabla_\mu T^\mn={F_2 \over f_2}\left(g^\mn \cl-T^\mn\right)\nabla_\mu R~~, \label{cov} \eeq

\noindent can be non-vanishing.

Using the projection operator

\beq P_\mn = u_\mu u_\nu + g_\mn \rightarrow P_\mn u^\mu = 0 ~~, \eeq

\noindent \eq{cov} yields

\beq(\rho + p) a_\mu = - P_\mu^\be p_{,\be} + {F_2 \over f_2} (u_\mu u^\al + \de_\mu^\al) (\cl - p ) R_{,\al}~~. \label{genEuler} \eeq

\noindent If $F_2 (\cl - p) \neq 0$, then the energy-momentum tensor is not covariantly conserved. As will be shown in subsequent sections, this possibility will play a crucial role in the dynamical behaviour of the pressure inside a homogeneous spherical body. Notwithstanding, below a general discussion of the implications of this non-conservation is presented.

The non-conservation expressed in \eq{cov} implies that test particles may deviate from geodesic motion: furthermore, since the {\it r.h.s.} of the above can depend on the constitution and structure of test bodies, free-fall may no longer be universal (see Refs. \cite{puetz1,puetz2,puetz3} for a general discussion) --- {\it i.e.} the WEP no longer holds (see Ref. \cite{OBJP2012} for an extensive discussion on this foundational principle of GR and its current experimental tests).

Indeed, \eq{cov} can be used to compute the force exerted upon a test particle,

\beq {d u^\alpha \over ds} + \Ga_\mn^\alpha u^\mu u^\nu = {1 \over \rho + p} \left[ {F_2 \over f_2} (\cl + p) R_{,\be} + p_{,\be} \right] P^{\alpha \be} \equiv f^\alpha ~~, \label{force} \eeq

\noindent clearly showing that, aside from the classical force due to the pressure of the fluid, an additional contribution due to the NMC arises.

In vacuum or a laboratory setting, if one simply assumes that the scalar curvature is negligibly small and sufficiently smooth, $R \sim R_{,\mu} \sim 0$, then the latter vanishes trivially and the WEP is recovered. However, one may resort to a more evolved scenario where the curvature is so low that the NMC is perturbative, $f_2(R) \sim 1$, and may be linearized as 

\beq f_2(R) \approx 1 + \be_2 {R \over \ka}  ~~, \label{linearcoupling} \eeq

\noindent with the curvature approximately given by its unperturbed expression $ R\approx \rho/2\ka $ (as further corrections would lead only to higher order terms).

In order to highlight the effect of the NMC contribution to the force \eq{force}, one studies a dust matter distribution with negligible pressure (so that the former dominates); one also considers a Lagrangian density $\cl = - \rho $ \cite{Lagrangian} (as the alternate choice, valid for scalar fields, is vanishingly small, $\cl = p \sim 0$).

For simplicity, one-dimensional motion in the $x$ direction with non-relativistic speed $v \ll 1 $ is finally assumed, so that the corresponding force is 

\beq f^x  \approx - {\be_2 \over 2\ka^2} (\rho' + \dot{\rho} v) ~~, \label{force1} \eeq

\noindent where $\rho' = d\rho/dx$ (or the analog expression with $x \leftrightarrow r$ for spherical symmetry).

In Ref. \cite{preheating}, it was found that a linear coupling between curvature and matter of the form of \eq{linearcoupling} is compatible with Starobinsky inflation and able to drive post-inflationary preheating if $ 10^{10} <\be_2 < 10^{14}$. Considering a lower bound $f^x > 10^{-13}~m/s^2$ (one order of magnitude below the precision of state of the art accelerometers), one obtains 

\beq  \left(\rho' + \dot{\rho} {v\over c} \right) \gtrsim {\ka^2 \over \be_2} {10^{-12} ~{\rm m / s^2} \over c} \sim {10^{89} \over \be_2 }~ {\rm kg.m^{-3}.s^{-1}} >  10^{79}~ {\rm kg.m^{-3}.s^{-1}} ~~. \label{force2} \eeq

\noindent One thus concludes that an observation of the breaking of the WEP due to the NMC requires an extremely high and unattainable density gradient $\rho'$, or an even more refined capability to manipulate $\dot{\rho}$.

Other tests of the effect of a NMC encompass searches for a putative ``fifth force'' perturbatively affecting orbital motion or E\"{o}tv\"{o}s-like experiments; indeed, assuming also that

\beq f_1(R) \approx R + \be_1 {R^2 \over \ka}  ~~, \eeq

\noindent one finds that a Yukawa contribution arises,
\beq U_Y \sim \alpha e^{-r/\lambda}~~,\eeq

\noindent with characteristic range $\la = \sqrt{\ka / \be_1} $ and strength $\alpha = \left( 1 - \be_2/\be_1 \right)/3$ \cite{NCB}; notice that a quadratic term for $f_1(R)$ is required to provide the range of the Yukawa interaction, as $\be_1 = 0 $ implies $\la \to \infty $ and $U_Y$ is then absorbed into the definition of the gravitational constant, $G = G_N(1+\al)$.

From the above, Ref. \cite{NCB} concludes that an NMC is compatible with current observational bounds on the strength and range of such Yukawa addition, provided that $\be_1 \sim \be_2$, as this yields $\alpha \lesssim 1$, well within the current experimental constraints, for instance, of sub-millimeter laboratory fifth force searches.

\section{Stationary case}

Imposing spherical symmetry and stationarity, one adopts the line element

\beq ds^2 = -e^{2\phi(r)} dt^2+ e^{2\la(r)} dr^2 + r^2 d\Om^2 ~~, \eeq

\noindent so that \eq{genEuler} becomes

\beq (\rho + p) \phi'(r) =  - p' + {F_2 \over f_2} (\cl - p) R'(r)~~. \label{genEulerstatic} \eeq

Introducing the mass function $m(r)$ through

\beq e^{-2\la} = 1 - { m(r) \over 8\pi \ka r}~~,  \eeq

\noindent the $0-0$ component of \eq{field0} becomes

\beqa && \label{field1-00}  m' \left({2 \over r}+  {d \over dr} \right) \left( \ka F_1 + F_2 \cl \right) =  8\pi \ka r  \left[ \ka (F_1 R - f_1 ) +  f_2 \rho + F_2 R \cl \right] + \\ \nonumber && \left[ 2\left(  8\pi \ka r - m \right) {d^2 \over dr^2} + \left( 32\pi \ka  - {3m \over r} \right) {d \over dr} \right] \left( \ka F_1 + F_2 \cl \right)  ~~, \eeqa

\noindent while the $r-r$ component reads

\beqa \label{field1-rr} &&  \phi' \left[ \ka F_1 + F_2 \cl + {r \over 2} {d \over dr} \left( \ka F_1 + F_2 \cl \right) \right]  = \\ \nonumber && \left[ { m \over 2r(8\pi \ka r - m)}  - {d \over dr} \right] \left( \ka F_1 + F_2 \cl \right) + { 2\pi \ka r^2 \over 8\pi \ka r - m} \left[ \ka ( f_1 - F_1 R) + f_2 p - F_2 R \cl \right] ~~. \eeqa

\noindent The trace of the equations of motion, \eq{trace0}, becomes

\beqa && \label{trace1} 3 \left( 1 - {m \over 8\ka r} \right) \left[ {d^2 \over dr^2}+ \left( \phi' + {3 \over 2r} + {1 \over 2}{ 8 \pi \ka - m' \over 8\pi \ka r - m } \right){d \over dr} \right]  \left( \ka F_1 + F_2 \cl \right)  = \\ \nonumber && {1 \over 2} f_2 (3p-\rho) +2\ka f_1 -\left( \ka F_1 + F_2  \cl \right) R ~~.\eeqa

One thus obtains three differential equations for four unknowns, $m(r)$, $\phi(r)$, $\rho(r)$ and $p(r)$. Solving these requires an additional equation, namely an EOS relating the pressure with the energy density, $p=p(\rho)$. In the following sections one assumes instead a homogeneous density, which allows for a considerable simplification of the dynamical behaviour of the aforementioned system ---  thus highlighting the impact of a NMC between matter and curvature and the choice of the Lagrangian density for matter.

Notice that in the context of models with a NMC between curvature and matter, black hole solutions can be obtained in the de Sitter background and it is found that the NMC ``dresses'' the cosmological term \cite{Bertolami:2014pga}; the same can be stated about charged black holes, where charges have to be suitably masked.

\subsection{Homogeneous sphere}

In order to isolate the effect of the NMC, one considers the linear form for the curvature term $f_1(R) = R$. One studies the impact of the former on a homogeneous sphere, $\rho = \rho_0 $: although this is not a realistic density profile, it yields a more tractable problem which allows one to determine how the pressure inside the body varies in order to counteract the gravitational pull of matter, and how a NMC affects the usual description of GR --- namely what is the relation between size and mass of a star above which gravitational collapse occurs. 

Considering the issue of how to properly choose the Lagrangian density of a perfect fluid, one writes $\cl = - \al \rho = - \al \be_0 \ka^2$, with $\al = 1$ or $\al = -\om(r)$ and $\be_0 \equiv \rho/\ka^2$ --- where $\om(r) = p(r) /\rho$ is the EOS parameter. With the above, \eq{genEulerstatic} becomes

\beqa (1 + \om) \rho_0 \phi' &=&  - \om' \rho_0 - {F_2 \over f_2} (\al + \om) \rho_0 R' \rightarrow \\ \nonumber -{\om' \over 1 + \om} &=& \phi'  + {F_2 \over f_2} {\al + \om \over 1 + \om} R' ~~. \label{genEulerstaticHomo} \eeqa

Noticing that the combination

\beq \ga \equiv {\al + \om \over 1 + \om} = \left\{\begin{matrix} 0~~~~,&~~~~\cl = p \rightarrow \al = -\om \\ 1~~~~,&~~~~ \cl = -\rho \rightarrow \al = 1 \end{matrix} \right. ~~, \eeq

\noindent acts as a ``binary'' variable, the above equation can be integrated,

\beq -{\om' \over 1 + \om} = \phi'  + \ga {F_2 \over f_2} R' \rightarrow \om = A{e^{-\phi} \over  f_2^{\ga}} - 1~~. \label{genEulerstaticHomo2} \eeq

If one considers a non-relativistic dust distribution with vanishing pressure, $\om=0$, and assumes that $\cl = -\rho \rightarrow \ga = 1$, the above can be recast as

\beq f_2^{\ga} \propto {1 \over  \sqrt{-g_{00}}}  ~~, \eeq

\noindent  a relation previously found in Ref. \cite{Martins}. One aims here to further explore the insight gained from that study, allowing for a non-vanishing pressure and the two possible choices of Lagrangian density already discussed.

Together with Eqs. (\ref{field1-00})-(\ref{trace1}) and the definition of the scalar curvature $R$, one has a closed set of equations for $\om$, $\phi$ and $m$,

\beqa \label{EE1} && m' \left({2 \over r}+  {d \over dr} \right) \left( F_2 \al - { 1 \over \be_0 \ka } \right) = \\ \nonumber && 8\pi \ka r  \left( F_2 R \al - f_2 \right) +\left[ 2\left ( 8\pi \ka r - m \right) {d^2 \over dr^2} + \left( 32\pi \ka  - {3m \over r} \right) {d \over dr} \right] \left( F_2 \al \right) ~~,\\ \nonumber && \left( {1 \over \be_0 \ka} + F_2  \al \right) R  = \\ \nonumber &&  {1 \over 2} f_2 (1 - 3\om) - 3 \left(1 - { m \over 8\pi \ka r}\right) \left[{d^2 \over dr^2}+ \left( \phi'+{3 \over 2r} + {1 \over 2}{ 8 \pi \ka- m' \over 8\pi \ka r - m } \right){d \over dr} \right] \left( F_2 \al \right)  ~~, \\ \nonumber && \om = A{e^{-\phi} \over  f_2^{\ga}} - 1~~, \eeqa

\noindent with the scalar curvature given by

\beq R = \sqrt{1-{m \over 8 \pi  r \ka }} \left[ {2 \over r^2} + \left({1 \over 2r}{m- m' r \over 8 \pi  r \ka - m }+ \phi'\right)\left({2\over r} + {1\over 2} \phi' \right) + \phi''  \right] - {2 \over r^2} ~~.\eeq

\section{Linear Coupling}

As before, one considers a linear coupling between curvature and matter

\beq f_2(R) = 1 + \be_2 {R \over \ka}~~,\eeq

\noindent as, following the discussion of section \ref{conservation}, this may be considered as a suitable approximation in the low curvature regime, and yields a more tractable problem that allows for the direct extraction of relevant consequences of the NMC; one also defines the dimensionless parameter $\ep \equiv \be_0\be_2 $. As expected, one finds that GR is recovered if either the coupling between matter and curvature vanishes,  $\be_2 = 0$, or if there is no matter, $\rho \sim \be_0 = 0$.

Considering that the density of the spherical body should not exceed the typical estimated value at the core neutron star, $\rho < \rho_N = 10^{18}~{\rm kg/m}^3$, and recalling the constraint $10^{10} <\be_2 < 10^{14}$ introduced after \eq{force1}, one gets

\beq \label{boundepsilon} \ep \equiv  \be_0\be_2 = {\be_2 \rho \over \ka^2}<  {\be_2 \rho_N \over \ka^2} \sim 10^{-62} ~~,\eeq

\noindent indicating that the NMC is highly perturbative.

The dimensionless functions are introduced below,

\beq \varrho \equiv {R \over 8\pi \ka \be_0}~~~~,~~~~\mu \equiv \sqrt{\be_0 \over 8\pi \ka}m ~~,\eeq

\noindent written in terms of the dimensionless variable $ x \equiv \sqrt{8\pi \ka \be_0} r$, so that \eq{EE1} becomes

\beqa \label{EE2} && \mu' \left({2 \over x}+  {d \over dx} \right) \left( \al -  {1 \over \ep} \right) = \\ \nonumber && - x \left( \varrho ( 1 - \al ) + { 1\over 8\pi \ep} \right) +  \left[ 2 \left ( x - \mu \right) {d^2 \over dx^2} + \left( 4  - {3 \mu \over x} \right) {d \over dx} \right] \al  ~~,\\ \nonumber &&\left( {1 \over \ep} + \al + {3\om - 1 \over 2} \right) \varrho = \\ \nonumber &&  {1 - 3\om \over 16\pi\ep} -  3 \left(1 - { \mu \over x}\right) \left[ {d^2 \over dx^2}+  \left( \phi'+ { 4x - 3\mu - \mu' x \over 2x(x - \mu) } \right){d \over dx} \right] \al ~~, \\ \nonumber && \phi' = { \ep \al' + {1 \over x - \mu} \left[ {x^2 \over 4} \left[ {\om \over 8\pi} + (\al + \om) \ep \varrho \right] + {\mu \over 2x} \left( 1 - \al \ep \right)  \right] \over 1 - \ep \left(\al + {1 \over 2} \al' x \right) } ~~,\\ \nonumber && \om =  A{e^{-\phi} \over  \left(1 + 8\pi \ep \varrho\right)^{\ga}} - 1~~,\eeqa

\noindent with the prime now denoting a derivative with respect to $x$.

\section{$\cl = -\rho$ case}

If one considers that $\cl = - \rho$ is the Lagrangian density of a perfect fluid, then the scenario of a homogeneous sphere naturally yields a very simplified set of equations: since both $F_2 $ and $\cl$ are constants, the additional terms found in \eq{EE2} involving spatial derivatives of the latter vanish.

Substituting $\al = 1$ into \eq{EE2}, one gets

\beqa \label{EEal1} &&   \mu' = { x^2 \over 16\pi \left( 1 -  \ep \right) } \rightarrow \mu = { x^3 \over 48\pi \left( 1 -  \ep \right) }~~,\\ \nonumber && \left( {1 \over \ep} + {1+3\om \over 2} \right) \varrho = {1 - 3\om \over 16\pi\ep} ~~, \\ \nonumber &&  \phi' = { {x^2 \over 4} \left[ {\om \over 8\pi} + (1 + \om) \ep \varrho \right] + {\mu \over 2x} \left( 1 - \ep \right) \over (x - \mu)(1 - \ep) }~~,\\ \nonumber && \om = A{e^{-\phi} \over 1 + 8\pi \ep\varrho} - 1~~.\eeqa

\noindent The second equation above may be used to write

\beq \left[ 2 + \ep(1+3\om ) \right] \varrho' = -3\om' \left({ 1 \over 8\pi} + \ep \varrho \right)~~.\eeq

Using \eq{genEulerstatic}, one obtains

\beqa &&\phi' =  - {\om' \over 1 + \om} - {8\pi\ep \over 1 + 8\pi \ep \varrho} {\al + \om\over 1 + \om} \varrho'  \rightarrow \\ \nonumber &&{x \over 48\pi(1 -\ep) -  x^2 } =  -2{1-\ep \over (1 + \om)(1+3\om + 2\ep)}  \om'  \rightarrow \\ \nonumber &&  \sqrt{48\pi(1-\ep) - x^2} =  A {  1+3\om + 2\ep  \over 1+\om } ~~.\eeqa

\noindent The integration constant $A$ may be determined from the boundary condition $\om(x_1)=0$, where $x_1$ signals the boundary of the spherical object, $x_1 \equiv \sqrt{8\pi \ka \be_0} r_1$, with $r_1$ the physical radius of the latter. One thus obtains

\beq A = {\sqrt{48\pi(1-\ep) - x_1^2} \over 1 + 2\ep } ~~.\eeq

Defining

\beq \label{defy} y \equiv \sqrt{1 - {x^2 \over 48\pi(1-\ep)}} ~~~~,~~~~ y_1 \equiv y(x_1)~~, \eeq

\noindent one then has

\beq \label{omal1} \om = {( y_1 - y )(1+2\ep ) \over  (1+2\ep) y -3 y_1}  ~~. \eeq

The central pressure is given by

\beq \label{omcal1} \om_c \equiv \om(x=0) = {[ y_1 - 1 ](1+2\ep ) \over (1+2\ep) -3 y_1 }  ~~,\eeq

\noindent and collapse is inevitable if it diverges, $ \om_c \rightarrow \infty $, leading to

\beq y_1 = {1+2\ep \over 3}\rightarrow r_1^2 = { 8 \over 3}{(1-\ep)^2 ( 2+\ep)\over \ka\be_0} ~~. \eeq

\noindent As expected, the standard result of GR,

\beq r_1^2 = { 16 \over 3 \ka\be_0} = { 16 \ka \over 3 \rho} \rightarrow {GM \over r_1c^2} = { 4\over 9 } ~~, \eeq

\noindent is obtained by setting $\ep = 0$.

From \eq{EEal1}, one gets

\beq\label{varrhoal1} \varrho = {2(1+2\ep) y - 3 y_1(1+\ep )  \over 8\pi (1-\ep) \left[ (1+2\ep) y - 3(1+\ep ) y_1 \right]} ~~, \eeq

\noindent and

\beq g_{00} = -e^{2\phi} = -A\left[ (1 + 8\pi \ep \varrho)(1+ \om)\right]^{-2} = -{B \over 4}\left( 3 -{1+2\ep \over 1+\ep } {y \over y_1} \right)^2~~. \eeq

Continuity with the Schwarzschild exterior metric

\beq \label{outermetric} ds^2_+ = -\left(1 - {M \over 8\pi\ka r}\right)dt^2 + {1 \over 1 - {M \over 8\pi\ka r}} dr^2 + r^2 (d\th^2 + \sin^2 \th d\phi^2) ~~,\eeq

\noindent at $y=y_1$ implies that $m(r_1 ) = M$ and 

\beq B = 4 y_1^2 \left( {1+\ep \over 2+\ep } \right)^2 ~~,\eeq

\noindent so that

\beq \label{metric00al1} g_{00} = - \left[ {3(1+\ep)y_1 - (1+2\ep) y \over 2+\ep } \right]^2 ~~.\eeq

\noindent Notice that this matching procedure is only valid if the Birkhoff theorem holds, as otherwise one should consider an exterior solution different from the Schwarzschild one: in the context of this work, the latter is indeed obeyed because one is considering a trivial $f_1(R) =R $ function \cite{JB,Capozziello:2011wg,Nzioki:2013lca,Henttunen2014110}, so that the coordinate-dependent approach followed here is sufficient. A more general approach with an arbitrary $f_1(R)$ should rely on the covariant formulation of the junction conditions, which yield any discontinuity in the extrinsic curvature $K$ when crossing over the boundary of the spherical object \cite{Deruelle:2007pt,Guarnizo:2010xr} (see also Ref. \cite{collapse} and references therein for a thorough discussion of this procedure in the context of gravitational collapse, and Ref. \cite{Bertolami:2014pga} for a study of the ensuing black hole).

From the above set of results, one finds that the strength of the NMC must obey $\ep < 1$, so that all quantities are well defined (in particular, so that the dimensionless coordinate $y$ is real and the sign of the $00$ component of the metric is correct). Given the stringent bound, \eq{boundepsilon}, this requirement is automatically fulfilled.

If one relaxes the compatibility with the preheating scenario discussed in Ref. \cite{preheating}, then a dominant, negative NMC, $\ep \rightarrow -\infty $ is not precluded: from Eqs. (\ref{EEal1}), (\ref{omal1}), (\ref{varrhoal1}) and (\ref{metric00al1}), one sees that taking this limiting case yields 

\beqa \mu &\approx& -{ x^3 \over 48\pi \ep } \sim 0~~,  \\ \om &\approx& { x_1^2 - x^2 \over 96\pi \ep} \sim 0~~, \\  \varrho &\approx& {1 \over 8\pi \ep} \sim 0 ~~, \\ g_{00} &\approx& -1 + { 3x_1^2 - 2x^2 \over 48\pi \ep} \sim -1~~. \eeqa

\noindent Thus, one finds that a dominant negative NMC effectively masks the presence of a spherical body; in order to prevent this unphysical result, the former must be perturbative, $\ep \ll 1$.

\section{$\cl = p$ case}

If one instead considers that $\cl = p$ is the suitable Lagrangian density to describe a perfect fluid, then a more involved set of equations is expected, since in this scenario the spatial derivative terms found in \eq{EE2} do not vanish.

Substituting $\al(r)= -\om(r)$ into \eq{EE2}, one gets

\beqa \label{EEalminusom1} && x \left[ \varrho ( 1 +\om ) + { 1\over 8\pi \ep} \right] = \\ \nonumber && 2 \left ( \mu - x \right) \om'' + \left( {3 \mu \over x} + \mu' - 4 \right) \om' + {2 \over x} \left(   {1 \over \ep} + \om \right) \mu' ~~,\\  \label{EEalminusom2} && \left( {1 \over \ep} + {\om - 1 \over 2} \right) \varrho =  \\ \nonumber &&{1 - 3\om \over 16\pi\ep} + 3 \left(1 - { \mu \over x}\right) \left[ \om''+  \left( \phi'+ { 4x - 3\mu - \mu' x \over 2x(x - \mu )} \right)\om' \right] ~~, \\  \label{EEalminusom3} && \phi' = { -\ep \om' + {1 \over x - \mu} \left[ {x^2 \om \over 32 \pi} + {\mu \over 2x} \left( 1 + \ep \om \right)  \right] \over 1 + \ep \left(\om + {1 \over 2} \om' x \right) }~~,\\  \label{EEalminusom4} && \om = A e^{-\phi}  - 1~~.\eeqa

Using the definition of the scalar curvature and the last of the above equations, one can write

\beq \varrho = { 4 x - 3\mu\over x^2 }  {\om'  \over 1+\om}+ {\mu' \over x} \left({2 \over x} - {\om'  \over 1+\om}  \right) + {2\over x} ( \mu - x ) {2(\om' )^2 - \om'' (1+\om)\over (1+\om)^2} ~~, \eeq

\noindent so that Eqs. (\ref{EEalminusom1})-(\ref{EEalminusom4}) can be reduced to

\beq \label{odeset} { x\over 16\pi } +  {\mu' \over x} \left( \ep - 1 \right) = 2 \left ( \mu - x \right) \ep \om'' + \left( {3 \mu \over x} + \mu' - 4 \right)  \ep \om' + 2 {x-\mu \over 1+\om} \ep (\om' )^2  ~~, \eeq

\noindent and

\beq 2( 1 - \ep ) \om' + \ep x ( \om')^2  = {1+\om \over \mu - x} \left[ {x^2 \om \over 16 \pi} + {\mu \over x} \left( 1 + \ep \om \right)  \right]~~. \eeq

Solving for $\mu$, one finally gets

\beq \label{odemu} \mu = {x^2 \over 16\pi} {( 1+\om ) x \om + 32\pi ( 1 - \ep ) \om' + 16\pi \ep x ( \om')^2 \over 2( 1 - \ep ) x\om' + \ep x^2 ( \om')^2 - ( 1+\om ) \left( 1 + \ep \om \right) }~~.  \eeq

Considering that the boundary of the spherical body is signaled by a vanishing pressure, $\om(x_1) = 0$, one has

\beq \mu(x_1) = {x_1^2 \over 16\pi} { 32\pi ( 1 - \ep ) \om' + 16\pi \ep x_1 ( \om')^2 \over 2( 1 - \ep ) x_1 \om' + \ep x_1^2 ( \om')^2 - 1 }~~. \eeq

\subsection{Constant solution}

One notes that a constant pressure solution is available: indeed, setting $\om = {\rm const.}$, Eqs. (\ref{EEalminusom1})-(\ref{EEalminusom4}) yields

\beqa \label{EEalminusomconst}  \om &=& {1 \over 2\ep - 3 }~~, \\ \nonumber \mu &=&  { x^3 \over 48\pi  \left( 1 - \ep \right) } ~~,\\ \nonumber \varrho &=& {1 \over 8\pi (1-\ep) } ~~,\\ \nonumber \phi&=& {1\over 2}\log \left(1 - { x_1^2 \over 48\pi \left( 1 - \ep \right) }\right)~~, \eeqa

\noindent where $\phi$ has been normalized following the previous procedure to match the metric at the boundary of the spherical body.

Notice that this solution cannot simultaneously yield a positive mass $\mu >0$ ($\ep < 1$) and pressure $\om \gtrsim 0$ (if $\ep > 3/2$). Furthermore, if the effect of the NMC is perturbative, $\ep \sim 0$, a spherical body with positive mass and curvature is obtained, but with negative pressure, $\om \approx -1/3$.

Conversely, a dominant NMC $|\ep| \gg 1 $ leads to a Minkowski space with $\om \sim \mu \sim \varrho \sim 0$ and $g_\mn \sim \eta_\mn$, as found in the previous section if $\cl = - \rho$. Again, this is unreasonable, and thus implies that the effect of the NMC should be perturbative, as supported by Ref. \cite{preheating}.

\subsection{Numerical solution}

One may substitute \eq{odemu} into \eq{odeset} and solve the ensuing second-order differential equation for $\om$. To do so, one ascertains the typical order of magnitude of $x_1$, assuming a perturbative NMC,

\beq x_1 = \sqrt{8\pi \ka \be_0} r_1 =  \sqrt{8\pi \rho \over \ka} r_1 \sim \sqrt{96\pi G M \over r_1 }  ~~,\eeq

\noindent which, considering that the classical upper bound $GM/r_1c^2 \lesssim 4/9$ remains approximately valid if the NMC is perturbative, that is

\beq x_1 \lesssim 8\sqrt{2\pi \over 3} \sim 10 ~~.\eeq

\begin{figure} 

\includegraphics[width= \linewidth]{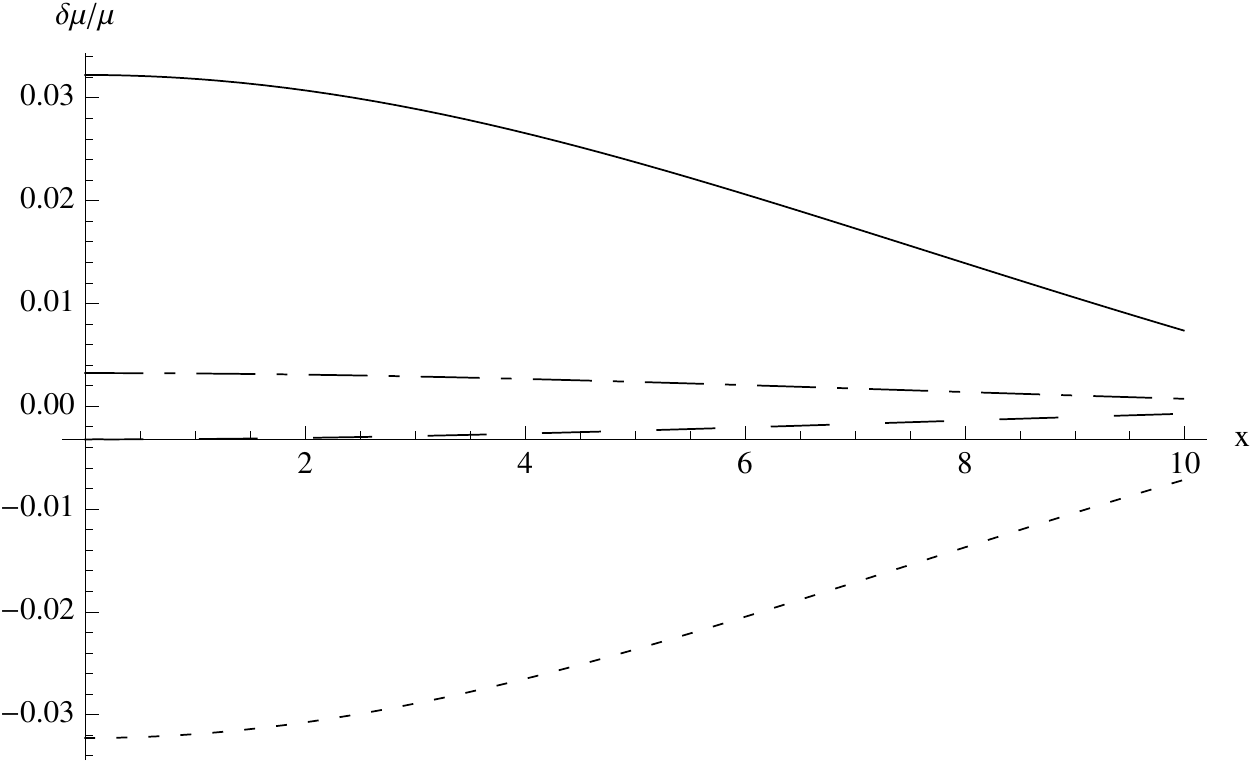}
\caption{Relative deviation of the dimensionless mass function $\mu$ for a spherical body with radius $x_1=10$, for $\ep = -10^{-3}$ (dashed), $-10^{-2}$ (dotted), $10^{-3}$ (dot-dashed) and $10^{-2}$ (full).}
\label{figrelmu}

\end{figure}

\begin{figure} 

\includegraphics[width= \linewidth]{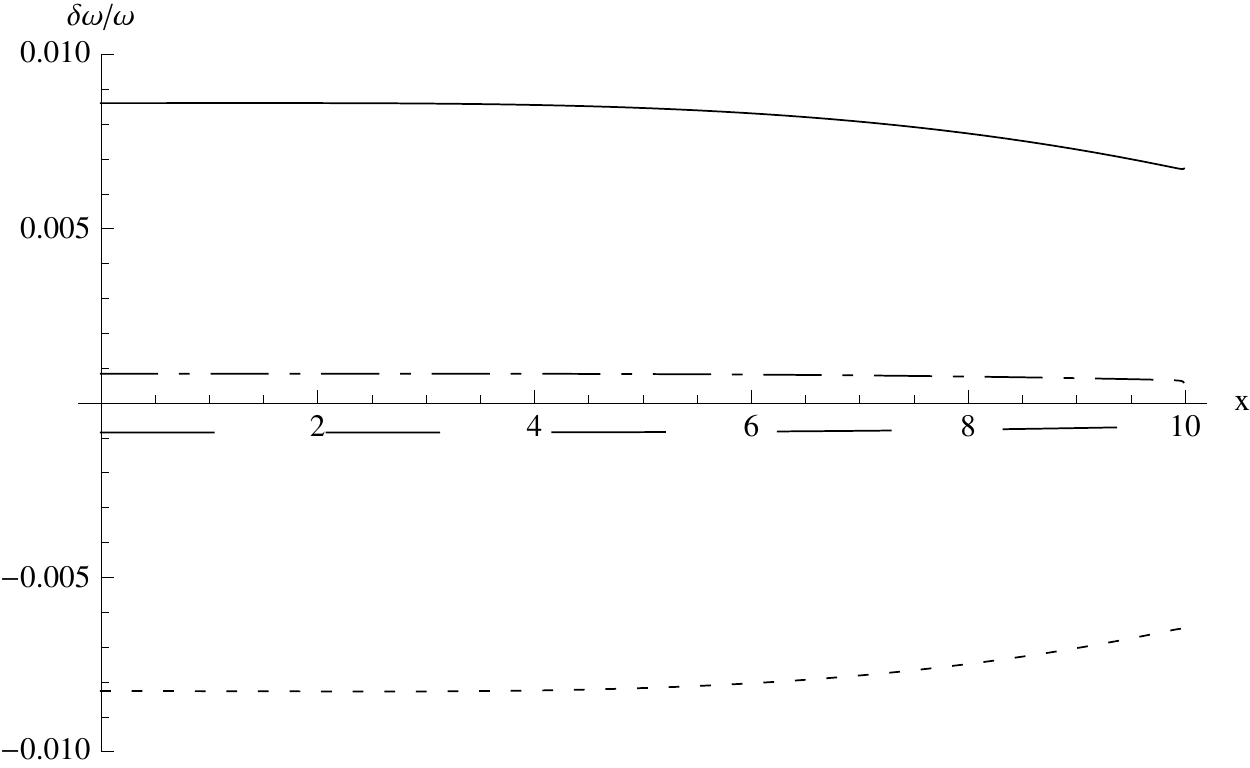}
\caption{Relative deviation of the EOS parameter $\om$ for a spherical body with radius $x_1=10$, for $\ep = -10^{-3}$ (dashed), $-10^{-2}$ (dotted), $10^{-3}$ (dot-dashed) and $10^{-2}$ (full).}\label{figrelom}

\end{figure}

Figs. \ref{figrelmu} and \ref{figrelom} show the numerical solution of Eqs. (\ref{odeset}) and (\ref{odemu}) for different values of the coupling strength $\ep$; boundary conditions $\om(x_1)=0$ and $\om'(0)=0$ are assumed, for $x_1 = 10$, the upper bound obtained above. The relative deviations $\de \mu/\mu \equiv 1 - \mu/\mu_{GR}$ and $\de \om/\om \equiv 1 - \om/\om_{GR}$ with respect to their GR counterparts $\mu_{GR}$ and $\om_{GR}$ are shown, with latter being defined as

\beq  \mu_{GR} \equiv { x^3 \over 48\pi}~~~~,~~~~ \om_{GR} \equiv { \sqrt{1 - {x_1^2 \over 48\pi}} - \sqrt{1 - {x^2 \over 48\pi}}  \over \sqrt{1 - {x^2 \over 48\pi}}  - 3 \sqrt{1 - {x_1^2 \over 48\pi}} } ~~. \eeq

Since the bound \eq{boundepsilon} for the latter indicates that it is almost vanishing, much higher values for $\ep$ are shown, in order to better illustrate the effect of the NMC.

%%%%%%%%%%%%%%%%%%%%%%%%%%%%%%%%%%%%%%%%%%%%%%%%%%%%%%%%%%%%%%%
%%%%%%%%%%%%%%%%%%%%%%%%%%%%%%%%%%%%%%%%%%%%%%%%%%%%%%%%%%%%%%%

%%%%%%%%%%%%%%%%%%%%%%%%%%%%%%%%%%%%%%%%%%%%%%%%%%%%%%%%%%%%%%%
%%%%%%%%%%%%%%%%%%%%%%%%%%%%%%%%%%%%%%%%%%%%%%%%%%%%%%%%%%%%%%%

%%%%%%%%%%%%%%%%%%%%%%%%%%%%%%%%%%%%%%%%%%%%%%%%%%%%%%%%%%%%%%%
\begin{figure} 

\includegraphics[width= \linewidth]{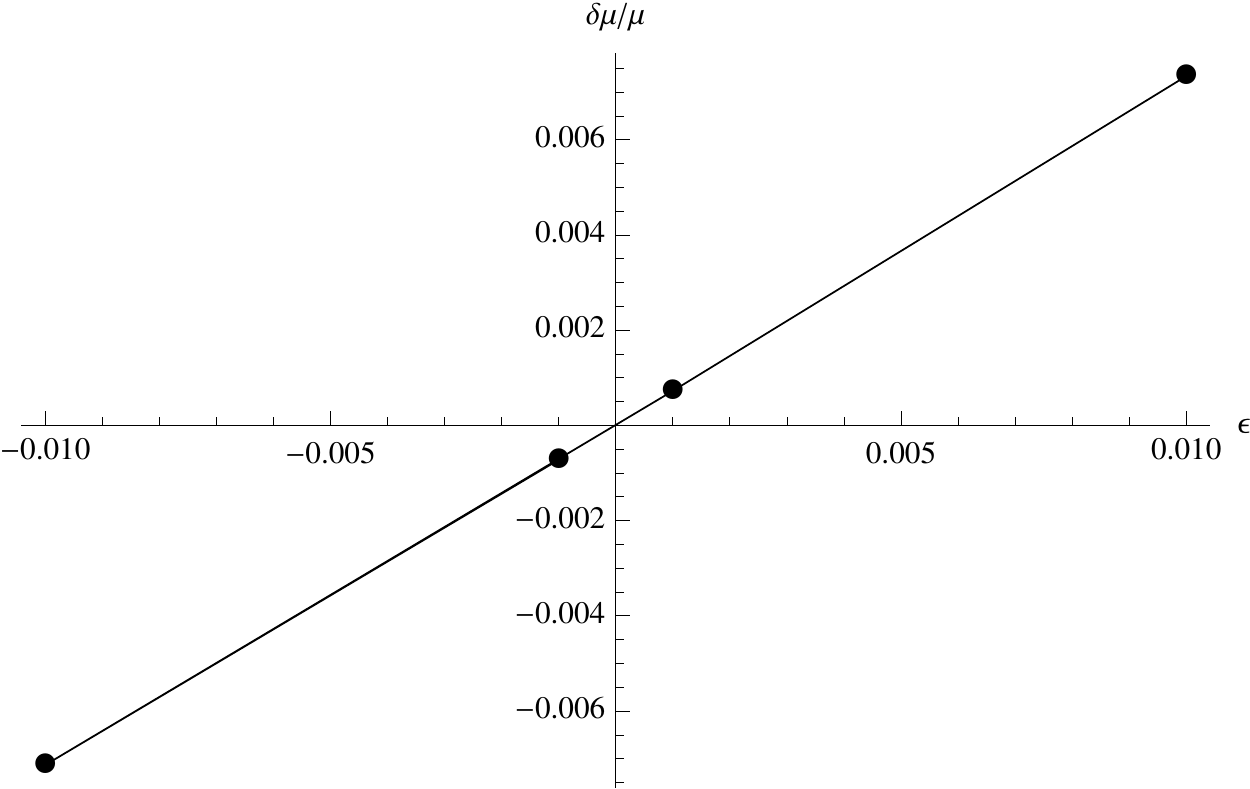}
\caption{Values of $1-\mu(x_1)/\mu_{GR}(x_1)$ as a function of $\ep$, with linear fit superimposed.}\label{linear}

\end{figure}
%%%%%%%%%%%%%%%%%%%%%%%%%%%%%%%%%%%%%%%%%%%%%%%%%%%%%%%%%%%%%%%

A numerical analysis does not yield an expression for the deviation of the Schwarzschild mass $M$ due to the effect of the NMC; since the latter is perturbative, one expects that the latter yields a linear correction to the GR value $M_{GR} = (4\pi/3)r_1^3\rho$; this is confirmed in Fig. (\ref{linear}), where the relative deviation is plotted together with a linear fit that allows one to estimate that

\beq 1 - {\mu(x_1) \over \mu_{GR}(x_1)} \sim {\de M \over M_{GR}} \sim 0.723 \xi ~~.\eeq

\section{Discussion and Outlook}

In this work, we have computed the effect of a linear coupling between matter and curvature on a spherical body with homogeneous density, for the choices of Lagrangian density $\cl = -\rho$ and $\cl= p$. In doing so, it complements two previous studies: one on the analogous effect on the Sun, modelled as a polytrope with polytropic index $n \sim 3$ \cite{solar}, and on the modification of the collapse of a homogeneous spherical body due to a linear NMC between curvature and matter \cite{collapse}.

Although the ensuing dynamical equations ruling the inner structure of such body are widely different, our results show that both formulations imply that the NMC should be perturbative, $f_2(R) \sim 1$ (consistent with previous studies): the converse would imply that the mass $M$ of the spherical body (as inferred from the Schwarzschild metric probed by an external observer) would be negative or, in the extreme case of a very large, negative NMC, vanish altogether --- and thus lead to an external Minkoswki spacetime, allowing for the masking of very large central masses.

Furthermore, since the widely different dynamical behaviour found in Eqs. (\ref{EEal1}) and (\ref{odeset})-(\ref{odemu}) is naturally suppressed by a perturbative NMC, our study shows that the effect of the latter on the mass $M$ is rather similar for both choices of Lagrangian densities,

\beqa \nonumber M = {4\pi \over 3}{\rho r_1^3 \over 1-\ep} \sim  {4\pi \over 3}\rho r_1^3 ( 1+\ep) ~~~~&,&~~~~\cl = -\rho ~~,\\ \nonumber M \sim   {4\pi \over 3}\rho r_1^3 ( 1+0.723\ep) ~~~~&,&~~~~\cl = p ~~.\eeqa

\noindent One highlights that the difference between the numerical factors is not only relatively small, but can be absorbed by the parameter $\ep = \be_0\be_2$, if one does not have {\it a priori} knowledge of the NMC strength $\be_2$; it could in principle be determined from the plethora of phenomena affected by a NMC (cited throughout this work). Alternatively, measuring the pressure profile inside the spherical body and comparing with the distinct expressions derived from $\cl = -\rho$ or $\cl = p$ would both allow the identification of the appropriate Lagrangian density and the determination of $\ep$.

Returning to the main motivation of this work, that is on the choice of the Lagrangian density in a non-minimally coupled model, one sees that it does not have a strong impact on the relevant observables: the results here presented indicate that in a stationary and perturbative regime, the selected form for $\cl$ does not affect greatly the impact on the structure of a spherical body. This is contrasting with respect with what occurs in a more dynamical context such as a gravitational collapse --- which, although beginning in a perturbative regime, inevitably evolves towards more extreme scenarios, with widely different consequences depending on the choices of $\cl$ \cite{collapse}.

This criterion allows us to reduce the degeneracy between different choices of NMC and Lagrangian densities: future studies aiming at testing the NMC should focus on perturbative, stationary scenarios. Conversely, we argue that if a NMC is assumed, the best environment to test what is the form of the Lagrangian density is found in time-evolving phenomena, where its effect eventually surfaces from an initial perturbative nature.

\begin{acknowledgements}
This work was partially supported by FCT (Funda\c{c}\~ao para a Ci\^encia e a Tecnologia, Portugal) under the project PTDC/FIS/111362/2009.
\end{acknowledgements}

\bibliographystyle{spphys}
\bibliography{homogeneous}{}

\end{document}